\documentclass[12pt,preprint]{aastex}

\newcommand{\etal}{\rm et~al.\ } 
\newcommand{\msp}{millisecond pulsar }
\newcommand{\msps}{millisecond pulsars }
\newcommand{\gp}{giant pulse }
\newcommand{\gps}{giant pulses }
\newcommand{\blc}{$B_{\rm LC}$ }
\newcommand{\pdot}{$\dot{P}$ }

\newcommand{\EAV}{$\langle E \rangle$ }
\newcommand{\eav}{\langle E \rangle }

\slugcomment{To be submitted to \apj}

\shorttitle{Giant Pulses from MSPs}
\shortauthors{Knight \etal}

\begin{document}

\title{A Search for Giant Pulses from Millisecond Pulsars}

\author{H. S. Knight\footnote{Affiliated with the Australia Telescope
National Facility, CSIRO}, M. Bailes}
\affil{Centre for Astrophysics \& Supercomputing, Swinburne University of Technology,
P.O. Box 218,Hawthorn VIC 3122, Australia}
\author{R. N. Manchester}
\affil{Australia Telescope National Facility, CSIRO, P.O. Box 76, Epping NSW 1710, Australia}
\and
\author{S. M. Ord}
\affil{Centre for Astrophysics \& Supercomputing, Swinburne University of Technology,
P.O. Box 218,Hawthorn VIC 3122, Australia}
\email{hknight@astro.swin.edu.au}

\begin{abstract}
We have searched for microsecond-timescale broadband emission from a
sample of eighteen millisecond pulsars.  Our study places strong
limits on such emission from several \msps and shows that it is only
present in a small subset of millisecond pulsars.  Giant pulses of up
to 64 times the mean pulse energy were detected from PSR J1823$-$3021A
in the globular cluster NGC 6624.  In contrast to the \gps of PSR
B1937+21, nearly all of the \gps from PSR J1823$-$3021A were
distributed within the trailing half of the main-pulse component of
the integrated pulse profile.  The fact that no \gps were observed on
the leading side of the main-pulse component suggests that \gps are
preferentially emitted closer to the last open field line than
ordinary emission.  The correlation between giant pulse emissivity and
spin-down luminosity in millisecond pulsars suggests that the high
period derivative of PSR J1823$-$3021A is intrinsic and is not just an
artifact of its acceleration in the gravitational potential of NGC
6624.
\end{abstract}


\section{Introduction}\label{sec:introduction}

The detection of strong bursts of radio emission emanating from the
Crab supernova remnant yielded the dramatic discovery of a central
neutron star in the form of the Crab pulsar \citep{sr68}.  These \gps
appear to be a distinct phenomenom to ordinary pulsed emission -- they
exhibit a variable spectral index consistent with or flatter than the
integrated profile \citep{sbh+99}, and a pulse energy distribution
showing a long high-energy tail with a power-law dependence on energy
\citep{lcu+95}.  The giant pulses are an intrinsically short timescale
phenomena with individual sub-pulses having durations as brief as two
nanoseconds \citep{hkwe03}.  Such ``nanopulses'' exhibit brightness
temperatures as high as $10^{37}$\,K.

\citet{cstt96} discovered \gps from a pulsar with substantially
different characteristics to the Crab pulsar.  Whilst the Crab pulsar
is just 950 years old, PSR B1937+21 has a characteristic age of 237
million years and spins over 20 times faster\footnote{Pulsar
parameters in this paper were obtained using the
\anchor{http://www.atnf.csiro.au/research/pulsar/psrcat}{ATNF Pulsar
Catalogue} \\ (http://www.atnf.csiro.au/research/pulsar/psrcat)}.  The
Crab pulsar's inferred surface dipole magnetic field is $10^4$ times
greater than that of PSR B1937+21.  Cognard \etal pointed out that the
two pulsars do however have the highest inferred magnetic fields at
their light cylinders ($B_{\rm LC}$) of all Galactic pulsars.
Discoveries of \gps from the high \blc pulsars J1824$-$2452
\citep{rj01} (hereafter RJ01) and B0540$-$69 \citep{jr03} have further
reinforced the notion that $B_{\rm LC} \propto P^{-2.5} \dot{P}^{0.5}$
is an indicator of \gp emissivity.  Despite the similarity in $B_{\rm
LC}$, \citet{kt00} showed that the \gp emission from PSR B1937+21
exhibits important differences to that of the Crab pulsar.  The \gps
from PSR B1937+21 are not coincident with the peak of the integrated
emission envelope, but instead are found in narrow $\sim 1\arcdeg$
phase windows on the extreme outer trailing edge of each of the main
and inter-pulse regions.  Interestingly, the pulsed X-ray flux from
PSR B1937+21 is coincident with the \gp emission window rather than
that of the ordinary emission \citep{chk+03,ncl+04}.

Few strong limits on \gp emission from \msps have been made due to
lack of instrumentation capable of searching data at high time
resolution.  \citet{es03a} and RJ01 searched for \gp emission from 5
and 11 millisecond pulsars respectively.  Edwards \& Stappers showed
\gp emission is not manifest in the \msp population as a whole.
However, it remains unclear how many \msps emit giant pulses of
durations much shorter than the sampling intervals of these studies.
In this study we obtain data on a large sample of bright \msps using a
baseband recorder system.  By implementing the coherent dedispersion
technique of \citet{hr75} we can remove dispersive smearing to the
nominal dispersion measure (DM) of the pulsar and obtain effective
sampling times $\sim$ 20--70 times shorter than Edwards \& Stappers
and RJ01.  Here we report new limits on \gp emission from several high
\blc pulsars, and investigate a previously unknown population of \gps
from PSR J1823$-$3021A.

\section{Observations and Data Analysis}\label{sec:observations}

Observations were taken using the Parkes 64-m radio telescope on 2004
February 1--4 and June 3--6.  Three observing bands centered at 685,
1341, and 1405 MHz were used.  Receivers used were the 10/50 cm
coaxial receiver for the 685 MHz observations and the H-OH receiver
for the 1341 and 1405 MHz observations.  These receivers have system
equivalent flux densities on cold sky of approximately 66 and 36 Jy
respectively.  All data were acquired and reduced using the CPSR2
baseband recorder \citep{bai03}.  This backend was used to 2-bit
sample at the Nyquist rate one or two dual-polarization 64 MHz bands.
The digitized data were subsequently distributed about an associated
cluster of 28 dual-processor 2.2 GHz Xeon computers.  Software then
formed a coherent filterbank of 256 channels dedispersed to the
nominal DM of the pulsar in a similar fashion to that detailed by
\citet{van03}.  After detection the synthetic filterbank data were
written to disk and total intensity pulse profiles were formed by
folding at the topocentric pulse period.

Data were searched at time resolutions in the range 4--128\,$\mu$s for
dispersed emission spikes.  Data segments of duration 16.7~s
containing consecutive samples totaling $11 \sigma$ above the local
mean were reduced in greater detail to produce plots of giant pulse
candidates.  The remaining data and their associated raw baseband
files that did not meet this criteria were deleted to free sufficient
disk space for further observations.

We applied our technique to short observations of PSR B1937+21 and PSR
J1824$-$2452 (PSR B1821$-$24) and quickly found giant pulses.
Subsequently we observed a sample of 18 millisecond pulsars, as
summarized in Table~\ref{tab:survey}.  Names and distinguishing
parameters of the pulsars (period, period derivative, spin-down
luminosity, and magnetic field at the light cylinder) are shown in
columns 1--5, the observation frequency is shown in column 6, while
the time and number of pulses observed are shown in columns 7--8.
Columns 9 and 10 show the mean pulse energy observed and the detection
energy threshold for emission lasting $\la 4$\,$\mu$s.  The effective
energy threshold is higher for pulses that are broader than 4\,$\mu$s.

\section{Results}\label{sec:results}

\subsection{PSR J1603$-$7202}\label{sec:1603}

Observations of the binary \msp J1603$-$7202 revealed 497
large-amplitude broadband spike detections.  The sample of pulse
events found were typically resolved to durations of 8--300\,$\mu$s.
This is in stark contrast to our observations of \gps from PSR
B1937+21 and PSR J1824$-$2452, where almost all \gps were unresolved
at the initial 4\,$\mu$s sampling time.  Many of the pulses from PSR
J1603$-$7202 had significant substructure across a wide phase range.
Although the pulses had very high peak flux densities, their energies
were typically only a few times the average pulse energy.  The large
pulse detection count is a result of our energy threshold being just
$0.2 \langle E \rangle$ at 4\,$\mu$s.  The emission events from PSR
J1603$-$7202 are therefore best interpreted as microstructure and are
not further discussed in this paper.

\subsection{PSR J1823$-$3021A}\label{sec:J1823$-$3021A}

PSR J1823$-$3021A is an enigmatic 5.4-ms pulsar that has the lowest
characteristic age, $\tau_{\rm char} = P/(2\dot{P}) = 25$ Myr, of all
known millisecond pulsars.  Its inferred spin-down energy loss rate
($\dot{E} \propto P^{-3}\dot{P}$) is the third highest of all \msps
after PSR B1937+21 and PSR J1824$-$2452.  However, PSR J1823$-$3021A
lies less than $1 \arcsec$ from the centre of the dense globular
cluster NGC 6624 \citep{sk95,hlk+04} and therefore acceleration in the
cluster potential undoubtedly contributes to the observed \pdot
\citep[see e.g.][]{sta97}.

\begin{figure}
  \begin{center}
    \includegraphics[scale=0.6,angle=270]{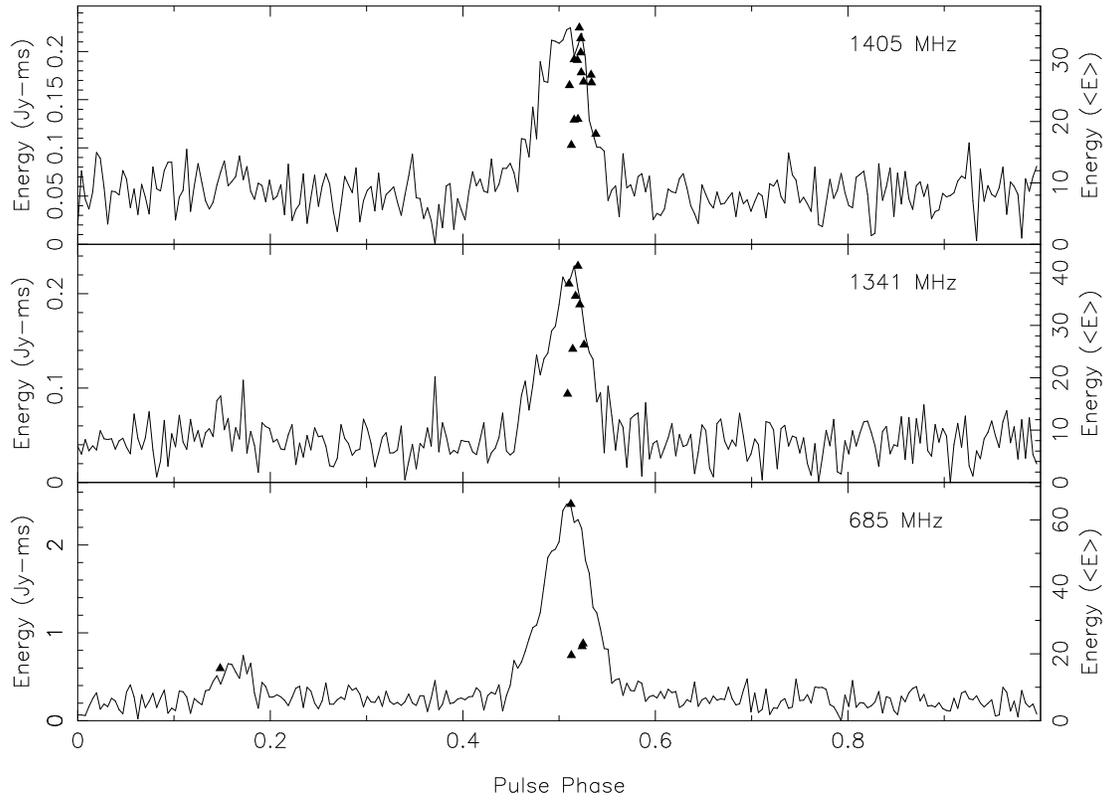}
    \caption{ Mean pulse profiles for PSR J1823$-$3021A at three
    frequencies. The triangles show the pulse phase and energy of
    the detected giant pulses.}
    \label{fig:phases}
  \end{center}
\end{figure}

In the 685, 1341, and 1405 MHz bands respectively, we detected 5, 7,
and 14 large-amplitude broadband emission events.  The signal to noise
ratios of the events were optimal at the DM of PSR J1823$-$3021A.  All
but one of the large-amplitude pulses were clustered around a small
phase window approximately 0.03 periods wide on the trailing half of
the dominant pulse component of PSR J1823$-$3021A (See
Fig.~\ref{fig:phases}).  The sole significant pulse found outside this
phase window was coincident with the earlier half of the small leading
component of PSR J1823$-$3021A.  The detection of this pulse at 685
MHz where only 5 of the 26 detections were made suggests the \gp
spectral index is steep within this window.  As all giant pulses found
are related in phase to emission from PSR J1823$-$3021A we conclude
PSR J1823$-$3021A is the source of the emission and not any other
cluster source.

\begin{figure}
  \begin{center}
    \includegraphics[scale=0.6,angle=270]{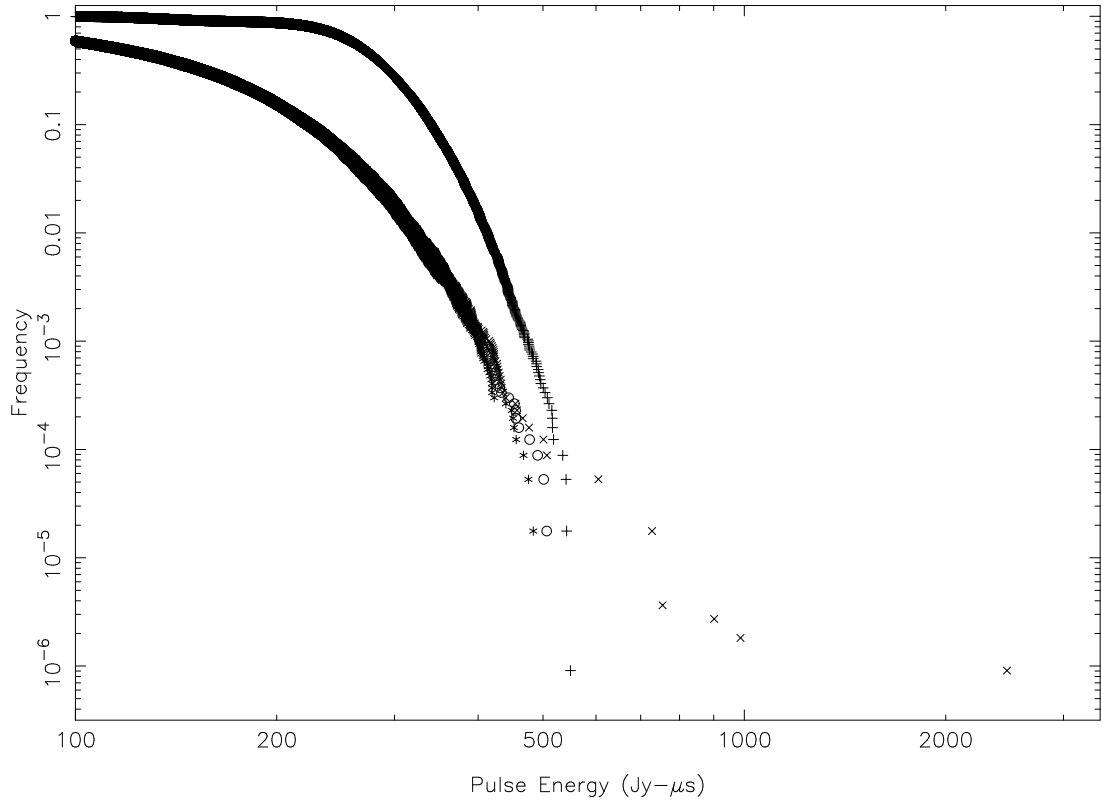}
    \caption{ Cumulative distribution of pulse energies at 685 MHz for PSR J1823$-$3021A.
    Stars denote a phase window
    [-0.04,0]; crosses denote [0,0.04]; circles denote [0.04,0.08]
    where phase zero is defined as the peak of the integrated profile.
    Pluses denote other pulsar phases.  }
    \label{fig:cumu}
  \end{center}
\end{figure}

Data not flagged as potentially containing a giant pulse were deleted
online.  Consequently only a few minutes of data were available
post-observation to determine the pulse energy distribution.  Each of
the remaining 685, 1341, and 1405 MHz datasets was re-folded to
produce single-pulse profiles of 128, 256, and 256 bins respectively.
By assuming each \gp to be just a single bin wide and obtaining the
peak flux for several phase windows the cumulative probability
distribution of pulse energy was determined.  The 685 MHz distribution
is shown in Fig.~\ref{fig:cumu}.  As the sample of pulses still
available is biased, the probabilities of the giant pulses that had
already been identified were shifted to take into account all data
taken.  At each frequency the pulses occurring on the second half of
the main pulse exhibit a clear power-law tail of giant pulses.
Fitting for the cumulative energy distribution yielded power-law
exponents of approximately $-2$ for the 685-MHz data and $-3$ for the
data at the two higher frequencies. While these indices appear
different, they are based on a small number of events; further
observations are required to verify any frequency dependence of the
power-law slope.

The mean FWHM of the \gps was 21\,$\mu$s for the 685 MHz pulses, and
7\,$\mu$s for the higher frequencies.  Visual inspection of the
stronger pulses at 685 MHz suggests that they are broadened by
interstellar scattering.  We attribute the greater detection counts at
1341 and 1405 MHz to their narrower pulse widths.  Future \gp searches
have the potential to be more productive at high frequencies where
scattering and dispersion will not dampen peak pulse fluxes as
markedly.  The dampening effect of pulse broadening can readily be
demonstrated by looking at the peak fluxes of the largest pulses.  At
685 and 1405 MHz the largest pulses peaked at 45 and 20 Jy, or around
680 and 1700 times the mean peak flux respectively.  The brightest
pulse at 1405 MHz would have a peak flux density of 8\,kJy at 1\,kpc
if it could be resolved to 1\,$\mu$s, which is less than the
corresponding values found for PSRs J1824$-$2452 (30\,kJy) and
B1937+21 (20\,kJy) \citep{jr03}.  When viewed in terms of the average
pulse energy the brightest pulses from PSR J1823$-$3021A in each band
had comparable energies (See Fig.~\ref{fig:phases}).  The most
energetic pulse was $64 \langle E \rangle$ at 685 MHz.  At 1405 MHz, 6
pulses were greater than 28 times the mean pulse energy, yielding $P(E
> 28\eav) \sim 4.6 \times 10^{-6}$.  In comparison, RJ01 measure for
PSR J1824$-$2452 at 1517.75 MHz $P(E > 28\eav) \sim 8.5 \times
10^{-7}$, which is around twice the value for PSR B1937+21.

\subsection{Limits for other Pulsars}\label{sec:others}

No large-amplitude emission events were detected for the remaining 16
millisecond pulsars.  Only PSRs J0034$-$0534, J1843$-$1113, and
B1957+20 had integrated flux densities lower than PSR J1823$-$3021A at
frequencies above 1300 MHz.  The null result for the majority of the
sources shows that \gp emission is not prevalent amongst millisecond
pulsars.  Of particular note were the lack of \gps from the
short-period and high \blc pulsars J1843$-$1113 and B1957+20.
\citet{jkl+04} report the detection from PSR B1957+20 of a single
large-amplitude pulse ($\sim 129 \eav$) and 5 marginal candidates in a
sample of $10^6$ pulses taken at 610 MHz.  It is difficult to
understand why Joshi \etal were able to detect large-amplitude
emission and we were not.  For PSR B1957+20 the central 48-MHz of our
band at 685 MHz gave approximately 1.2 times the sensitivity of the
16-MHz system of Joshi \etal for an equal sampling interval.
Furthermore, we observed nearly five times as many pulses from PSR
B1957+20 with an effective sampling rate that was over 64 times
faster.  One possible explanation is that PSR B1957+20 emits strong
\gps at a very low frequency and Joshi \etal were fortunate enough to
chance upon such an event, whilst we were not.  Another explanation is
that the solitary pulse of Joshi \etal is spurious.  Our observations
indicate that PSR B1957+20 should not currently be classed as a
millisecond pulsar which emits frequent or strong giant pulses.

\section{Discussion}\label{discussion}

The \gps from PSR J1823$-$3021A differ from those of PSR B1937+21 in
that they are not clustered around the extreme trailing edge of the
pulse components.  Such a difference is probably due to viewing
geometry rather than a different emission mechanism, and so a
geometric interpretation can be conjectured.  The two pulse components
of PSR B1937+21 are separated in phase by nearly $180\degr$ and
therefore the pulsar has been interpreted as an orthogonal rotator
\citep{stc99}.  With each pulse component having an associated \gp
emission region it is natural then to assume that both poles are
endowed with the conditions necessary for \gp emission.  Rather than
interpreting the two pulse components of PSR J1823$-$3021A as emission
from opposing poles, we suggest that they represent two cuts of a cone
of emission that lies close to the last open field line.  The \gps
appear to be preferentially emitted towards the outside edge of this
emission cone compared to the ordinary emission.  In addition, the
\gps may originate higher in the magnetosphere than ordinary emission.
Of course, with only one \gp detected on the leading component of PSR
J1823$-$3021A more data needs to be taken to clarify such speculation.
PSR B1937+21 is then interpreted as having narrow \gp envelopes much
closer to the last open field line than its ordinary emission
envelopes.

With around $\sim 70$\% of \msps found in binary systems it is
interesting to note that the three \msps now known to emit \gps are
all solitary.  Along with PSR J0737$-$3039A and the solitary globular
cluster pulsar J1910$-$5959D the three \msps in question have the
lowest characteristic ages of all millisecond pulsars.  If they were
spun-up through accretion from a binary companion \citep[see
e.g.][]{bv91} they must have ablated or lost their companions in their
short lifetimes.  If pulsars can ablate a binary companion in just a
few million years it is not surprising that it is the energetic
giant-pulse emitters that are solitary.  Observations of solitary
\msps in globular clusters could therefore yield further detections of
giant pulse emission.

Giant pulses of energy $\ga 28 \langle E \rangle$ emanating from PSR
J1823$-$3021A were found at a higher rate than for PSR B1937+21 and
PSR J1824$-$2452.  The rate at 1341 and 1405 MHz is sufficiently high
that if the pulsar were twice as weak our observations would only have
been detected it through giant pulses.  Despite the high rate of giant
pulse activity from PSR J1823$-$3021A we failed to detect any giant
pulse activity from other pulsars with similar or higher inferred
$B_{\rm LC}$ such as PSR B1957+20.  Radiation from PSR B1957+20
ablates gas from its companion \citep{fbb+90}.  Our non-detection of
\gps from PSR B1957+20 may result from the ablated gas causing excess
scattering on microsecond timescales.  Other systems like PSR
J1843$-$1113 and PSR J0034$-$0534 may not emit enough strong \gps for
us to detect or simply not emit giant pulses at all.

Acceleration of PSR J1823$-$3021A in the potential of NGC 6624 could
also mean that its intrinsic spin-down rate and \blc are greater than
the values inferred from observation.  \citet{bzr04} have shown that
PSR J1823$-$3021A exhibits cubic timing residuals and has a
spin-frequency second derivative $\ddot{\nu} = 6.1 \times
10^{-25}$\,s$^{-3}$.  The magnitude of $\ddot{\nu}$ is consistent with
the empirical relation for timing noise found by \citet{antt94}.  Such
timing noise is supportive of the suggestion that the pulsar's
intrinsic spin-down rate is large and that it is reasonably youthful.
However, explaining the high \gp emissivity of PSR J1823$-$3021A
through a large intrinsic $\dot{P}$ implies that the observed
characteristic age is larger than the true age.  Considering the
average characteristic age of field millisecond pulsars is 7 Gyr, the
chances of a millisecond pulsar with an age of $\la$25 Myr being
amongst the $\sim$100 known seem low.  Two unrecycled pulsars are also
found in NGC 6624 \citep{bbl+94,cha03}.  If PSR J1823$-$3021A does
have a low age then the existence of these short-lived slow pulsars
implies NGC 6624 could have a large pulsar formation rate in the
current epoch.  Further evidence favouring a low age for PSR
J1823$-$3021A could be made by observations of microglitches similar
to those observed in PSR J1824$-$2452 \citep{cb04}.

Phenomena such as field-line current sweepback mean that the magnetic
field at the light cylinder has a significantly different structure to
that of the surface field.  Despite this, we may conjecture that
magnetic inclination angle plays a role in explaining why PSR
J1823$-$3021A emits frequent \gps and why PSR B1957+20 has not been
found to do so.  However, \citet{jkl+04} report that PSR J0218+4232
emits large-amplitude pulses of up to 51 times the mean pulse
intensity.  PSR J0218+4232 has a larger \blc than PSR J1823$-$3021A,
but has significant unpulsed radio emission and has been interpreted
as a nearly aligned rotator \citep{nbf+95,stc99}.  If the
large-amplitude pulses of PSR J0218+4232 are giant pulses they spoil a
simple geometrical interpretation for why PSR B1957+20 does not appear
to emit strong giant pulses despite its high $B_{\rm LC}$.

The fact that the \gp emitting pulsars all have very high \blc is
unsurprising in many respects -- it is natural to expect that those
pulsars with the largest magnetic fields and fastest spin rates would
exhibit energetic phenomena not found in the emission of less extreme
millisecond pulsars.  Perhaps then there are other contributing
factors that are more important than $B_{\rm LC}$?  RJ01 have
suggested that a narrow X-ray pulse with a hard spectrum could be a
better indicator of a magnetospheric geometry conducive to \gp
emission.  Of the \msps observed in this study only PSR J2124$-$3358
has been found to pulse in the X-ray band \citep{bt99}.  However, its
pulse is not narrow like those of PSR B1937+21 and PSR J1824$-$2452
\citep{chk+03,ncl+04,mcm+04}.  X-ray timing observations of PSR
J1823$-$3021A could therefore help confirm whether or not the
existence of a narrow X-ray pulse correlates with emission of radio
giant pulses.  Spin-down energy and X-ray luminosity are linearly
correlated for rotation powered pulsars \citep{bt97}.  Therefore it is
interesting to note that PSRs B1937+21, J1824$-$2452, and
J1823$-$3021A have by far the highest inferred spin-down energy loss
rates of all millisecond pulsars. Perhaps then, at least for
millisecond pulsars, X-ray luminosity is a proxy and $\dot{E}$
dictates \gp emissivity.  Information on correlations with high energy
emission can be gleaned through further radio observations of PSR
J0218+4232, whose X-ray and $\gamma$-ray emission properties are
comparatively well known \citep{khv+00,khv+02,wob04}.

\section{Conclusions}\label{conclusions}

We have undertaken a survey for \gp emission from \msps and detected
strong microsecond-timescale \gps from PSR J1823$-$3021A.  The
strongest \gp was found at 685 MHz and had over 64 times the energy of
the average pulse.  The majority of \gps were found above 1300 MHz
where scattering was not significant at our 4\,$\mu$s sampling time.
No \gp emission was discovered from other pulsars with similar or
higher $B_{\rm LC}$ to that of PSR J1823$-$3021A.  In particular, no
giant pulses were found from B1957+20.  If the inferred \blc is the
sole determinant of \gp emissivity, the emission of \gps from PSR
J1823$-$3021A can be explained by invoking a cluster acceleration that
provides a large negative contribution to the measured $\dot{P}$ of
PSR J1823$-$3021A and consequently lessens the derived $B_{\rm LC}$.
This in turn would imply that PSR J1823$-$3021A is very young, and
that NGC 6624 presently has a large pulsar formation current.
Alternatively, ablated gas from the companion of PSR B1957+20 might
scatter-broaden any giant pulses below our detection threshold.  The
three \msps that are known to emit \gps have the highest inferred
spin-down luminosities of all millisecond pulsars, and therefore
$\dot{E} \propto P^{-3}\dot{P}$ rather than $B_{\rm LC} \propto
P^{-2.5} \dot{P}^{0.5}$ may be a better indicator of \gp emission.
Future studies of \gp emission at high radio frequencies will avoid
complications arising from scattering-induced signal degradation, and
therefore reveal the true widths and brightnesses of \gps and help
establish the true indicators of \gp emissivity.  X-ray timing
observations of PSR J1823$-$3021A offer the opportunity to see whether
there is a phase correlation between the \gp and X-ray emission
envelopes.  Such a correlation would add strong support to the notion
that \gps emanate from a similar region of the pulsar magnetosphere as
high energy emission.

\acknowledgements

The Parkes radio telescope is part of the Australia Telescope which is
funded by the Commonwealth of Australia for operation as a National
Facility managed by CSIRO.  We thank A. Hotan for observing
assistance, C. West for use of disks and W. van Straten for software
assistance.  HSK acknowledges the support of a CSIRO Postgraduate
Student Research Scholarship.

\clearpage

%\bibliographystyle{apj}
%\bibliography{journals,modrefs,psrrefs,crossrefs,hkrefs}

\clearpage

\begin{deluxetable}{lrcccccccc}
\tabletypesize{\scriptsize}
\tablecolumns{10} 
\tablewidth{0pc} 
\tablecaption{Pulsars searched for \gp emission.\label{tab:survey}} 
\tablehead{ 

\colhead{PSR}&
\colhead{$P$} &
\colhead{\pdot} &
\colhead{$\dot{E}$} &
\colhead{\blc} &
\colhead{$\nu$} &
\colhead{$t_{\rm obs}$} &
\colhead{$N_{\rm p}$} &
\colhead{\EAV} &
\colhead{$E_{\rm lim}$} \\
 &
\colhead{(ms)} &
\colhead{($10^{-21}$s/s)} &
\colhead{($\times 10^{32}$ ergs/s)} &
\colhead{($\times 10^3$G)} &
\colhead{(MHz)} &
\colhead{(s)} &
\colhead{($\times 10^5$)} &
\colhead{(${\rm Jy}\cdot{\rm \mu s}$)} &
\colhead{($\langle E \rangle$)} \\

}
\startdata 

% Pulsar   & P   & Pdot & Edot &B_LC  & Freq & Seconds    & Pulses& & \EAV& $E_{lim}$
J0034$-$0534& 1.88&  4.96& 300  &   138&   685& 2473,3251  & 13,17 & 4.2,6.3 & 39,26 \\
           &     &      &      &      &  1341& 12897      & 69    & not detected & unknown \\
J0613$-$0200& 3.06&  9.57& 130  &  56.6&  1341&  4056,2471 & 13,8.0& 19,13   & 4.5,6.3 \\
           &     &      &      &      &  1405&  3190      & 10    & 16      & 5.3 \\
J0711$-$6830& 5.49& 14.90& 36   &  16.4&   685&  1938      & 3.5   & 55      & 3.0 \\
           &     &      &      &      &  1341&  7874,2466 & 14,4.5& 21,36   & 3.9,2.3 \\
           &     &      &      &      &  1405&  7968      & 15    & 36      & 2.3 \\
J0737$-$3039A\tablenotemark{a}&22.70& 1740& 59 &5.10& 685& 4116       & 1.8   & 310     & 0.6 \\
           &     &      &      &      &  1341&  7775      & 3.4   & 100     & 0.8 \\
J1022+1001 &16.45& 43.41&   3.8&  1.80& 1341&  4772,4202 &2.9,2.5& 30,290  & 2.8,0.3 \\
           &     &      &      &      & 1405&  4951      & 3.0   & 19      & 4.3 \\
J1603$-$7202&14.84& 15.64& 1.9  &  1.40& 1341&  2315      & 1.6   & 360     & 0.2 \\
           &     &      &      &      & 1405&  2473      & 1.7   & 200     & 0.4 \\
J1629$-$6902& 6.00&  10.0& 18   &  10.8& 685 &  3652      & 6.1   & 61      & 3.0 \\
           &     &      &      &      & 1341&  2675      & 4.5   & 11      & 8.1 \\
           &     &      &      &      & 1405&  2655      & 4.4   & 11      & 7.7 \\
J1643$-$1224& 4.62& 18.49& 74   &  28.1& 1341&  2877      & 6.2   & 39      & 2.2 \\
           &     &      &      &      & 1405&  2874      & 6.2   & 36      & 2.4 \\
J1713+0747 & 4.57&  8.54& 35   &  19.6& 1341&  902       & 2.0   & 82      & 1.1 \\
           &     &      &      &      & 1405&  947       & 2.1   & 28      & 3.2 \\
J1730$-$2304& 8.12& 20.21& 15   &  7.17& 1341&  2659      & 3.3   & 51      & 1.9 \\
           &     &      &      &      & 1405&  2603      & 3.2   & 50      & 1.9 \\
J1732$-$5049& 5.31&  13.8& 36   &  17.1& 685 &  1493      & 2.8   & 55      & 4.1 \\
           &     &      &      &      & 1341&  3205      & 6.0   & 16      & 5.8 \\
           &     &      &      &      & 1405&  3204      & 6.0   & 12      & 7.4 \\
J1744$-$1134& 4.07&  8.94& 52   &  26.8& 1341&  2919      & 7.2   & 6.9     & 13 \\
           &     &      &      &      & 1405&  2558      & 6.3   & 15      & 6.0 \\
J1823$-$3021A&5.44&3384.14&8300\tablenotemark{b}&253\tablenotemark{b}& 685 &  6136      & 11    & 38    & 5.9 \\
           &     &      &      &      & 1341&  6029      & 11    & 5.4   & 17 \\
           &     &      &      &      & 1405&  7274      & 13    & 6.5   & 14 \\
J1843$-$1113& 1.85&  9.59& 600  &   201& 685 &  3160      & 17    & 11    & 25 \\
           &     &      &      &      & 1341&  8971      & 48    & 2.4   & 41 \\
J1857+0943 & 5.36& 17.84& 46   &  19.0& 1341&  3360      & 6.3   & 17    & 5.8 \\
           &     &      &      &      & 1405&  3355      & 6.3   & 53    & 1.8 \\
B1957+20   & 1.61& 16.85& 1600 &   376& 685 &  3385,4348 & 21,27 & 16,12 & 12,16 \\
           &     &      &      &      & 1341& 11009      & 68    & not detected & unknown \\
J2124$-$3358& 4.93& 20.54& 68   &  25.2& 1341&  1575      & 3.2   & 37  & 2.3 \\
           &     &      &      &      & 1405&  1580      & 3.2   & 39  & 2.1 \\
J2145$-$0750&16.05& 29.86& 2.8  &  1.59& 1341&  1340      & 0.83  & 50  & 1.7 \\
           &     &      &      &      & 1405&  1333      & 0.83  & 83  & 1.0 \\
\enddata
\tablenotetext{a}{Double pulsar.}
\tablenotetext{b}{Disregards acceleration in potential of globular cluster NGC 6624.}
\end{deluxetable}

\clearpage

\clearpage

\clearpage

\end{document}